\documentstyle[12pt]{article}

\documentstyle{titlepage}
\title{SOME OBSERVABLE RESULTS OF THE RETARDED BOHM'S THEORY}
\author{\bf ALI\ SHOJAI$^*$\ \&\ MEHDI\ GOLSHANI$^{**}$\\
Department of Physics, Sharif University of Technology\\P.O.Box 11365-9161 Tehran, IRAN\\
and\\Institute for Studies in Theoretical Physics and Mathematics,\\P.O.Box 19395-5531, Tehran, IRAN\\
$^*$Email: SHOJAI@PHYSICS.IPM.AC.IR\\
$^{**}$Fax: 98-21-8036317\\}
\date{}
\begin{document}
\begin{bf}
\maketitle
\vspace{1cm}
\begin{center}
{\Large SOME OBSERVABLE RESULTS OF THE RETARDED BOHM'S THEORY}\\
{\bf A. Shojai \& M. Golshani}
\end{center}
\vspace{0.5cm}
\begin{center}
{\bf ABSTRACT}
\end{center}
{\it It is shown that the retarded Bohm's theory has at l
east four novel
properties. (1) The center of mass of an isolated two-body system is
accelerated. (2) Hydrogen-like atoms are unstable. (3) The distribution function
differs from the standard one. (4) The definition of energy needs some care.}
\vspace{1.5cm}
\section{INTRODUCTION}
\par
Nonrelativistic Bohm's theory (NBT) can be formulated in terms of the following three
postulates[1]:\\
(I)--- Any system of particles is always accompanied by an objectively real
field ($\psi (\vec{x}_1\cdots\vec{x}_N;t)$), which satisfies the Schr\"odinger
equation:
\begin{equation}
i\hbar\frac{\partial \psi}{\partial t}=\sum_{i=1}^N-\frac{\hbar^2}{2m_i}\nabla^2_i\psi+V(\vec{x}_1\cdots\vec{x}_N;t)\psi
\end{equation}
(II)--- Particles move according to:
\begin{equation}
\frac{d\vec{r}_i(t)}{dt}=-i\frac{\hbar}{m_i}\left [ \vec{\nabla}_i\left \{\ln \left (\frac{\psi}{\sqrt{\psi^*\psi}}\right ) \right \} \right ]_{\{\vec{x}_j\}=\{\vec{r}_j(t)\}}
\end{equation}
(III)--- The distribution function of an ensemble of such system is given by:
\begin{equation}
\rho(\vec{x}_1\cdots\vec{x}_N;t)=\psi^*\psi
\end{equation}
\par
It can be shown that the second and third postulates are compatible. That is, 
the motion predicted by (II) preserves the distribution function given by (III)[1,2].
\par
A simple-minded extension of this formalism to the relativistic domain (i.e.
simply writing the Lorentz covariant analoges of (1)--(3)) fails[2,3]. This is mainly  
because NBT is highly non-local.

\par
Recently a local relativistic Bohm's theory, called retarded Bohm's
theory (RBT)[4], is introduced. It is basically founded on the assumption 
that for the calculation of the position of some particle, the position of others should be evaluated at 
the retarded times. This means that instead of (2) we must use:
\begin{equation}
\frac{d\vec{r}_i(t)}{dt}=-i\frac{\hbar}{m_i}\left [ \vec{\nabla}_i\left \{\ln \left (\frac{\psi}{\sqrt{\psi^*\psi}}\right ) \right \} \right ]_{\{\vec{x}_j\}=\{\vec{r}_j
(t_{ij})\}}
\end{equation}
where $t_{ij}$ is the retarded time of the $j$th particle with respect to the
$i$th particle, defined by:
\begin{equation}
t_{ij}=t-\frac{|\vec{r}_i(t)-\vec{r}_j(t_{ij})|}{c}
\end{equation}
This formalism is manifestly relativistic, in the sense that actions are propagated by light's velocity.
\par
There are at least three questions about this theory.
The first oneconcerns the problem of whether it has any prediction beyond
the standard quantum mechanics which can be
 checked experimentally? The inventor of RBT and others have presented some
samples of such experiments[4,5]. In this work we shall present two other ones. These involve the self-acceleration
of the center of mass of an isolated two-body system, and some sort of
unstability in Hydrogen-like atoms.
\par
The second question is about the consistensy between (3) and (4). Finally the third question is related 
to the problem of defining the energy of a system in RBT. These questions 
are also investiga
ted in this paper.
\section{Self-acceleration Effect}
\par
In NBT as in the classical mechanics the center of mass of an isolated system is not accelerated. 
For an isolated two-body system we have:
\begin{equation}
V(\vec{x}_1,\vec{x}_2;t)=V(|\vec{x}_1-\vec{x}_2|)
\end{equation}
It can be easily shown that the solution of Schr\"odinger equation is:
\begin{equation}
\psi(\vec{x}_1,\vec{x}_2;t)=\Phi (\vec{x}_1-\vec{x}_2)e^{i\vec{K}\cdot \vec{X}_{c.m.}}e^{-iEt/\hbar}
\end{equation}
where:
\begin{equation}
\vec{X}_{c.m.}=\frac{m_1\vec{x}_1+m_2\vec{x}_2}{m_1+m_2}
\end{equation}
The center of mass velocity is:
\begin{equation}
\vec{v}_{c.m.}=\frac{\hbar \vec{K}}{m_1+m_2}=constant
\end{equation}
where we have used the fact that:
\begin{equation}
\vec{\nabla}_1\Phi=-\vec{\nabla}_2\Phi
\end{equation}
\par
On the other hand, in RBT the right-hand-side of (4) is not calculated at the same 
time for the two particles, and since there is in general
an asymmetry between $t_{12}$ and $t_{21}$, 

as can be seen from the Fig. (1),
the center of mass velocity would not
be a constant.
\par
In order to clarify this point, let us study a Hydrogen-like atom with $\vec{K}=0$ and:
\begin{equation}
\Phi=R_{nl}(r)P_{l1}(\theta)e^{i\phi}
\end{equation}
where $r$, $\theta$ and $\phi$ are the spherical coordinates of the relative 
distance $\vec{x}=\vec{x}_1-\vec{x}_2$. It can be shown easily that:
\begin{equation}
t_{12}=t-\frac{a}{c}
\end{equation}
\begin{equation}
t_{21}=t-\frac{\tilde{a}}{c}
\end{equation}
where:
\begin{equation}
a=r+\frac{1}{c}(x\dot{x}_2+y\dot{y}_2)+{\cal O}(v^2/c^2)
\end{equation}
\begin{equation}
\tilde{a}=r-\frac{1}{c}(x\dot{x}_1+y\dot{y}_1)+{\cal O}(v^2/c^2)
\end{equation}
The equations of motion (4) lead to the following relations:
\begin{equation}
m_1\dot{x}_1= -\frac{\hbar}{r^2}\left [ y-\frac{2y}{rc}(x\dot{x}_2+y\dot{y}_2)+\frac{r}{c}\dot{y}_2\right ]+{\cal O}(v^2/c^2)
\end{equation}
\begin{equation}
m_2\dot{x}_2= +\frac{\hbar}{r^2}\left [ y+\frac{2y}{rc}(
x\dot{x}_1+y\dot{y}_1)-\frac{r}{c}\dot{y}_1\right ]+{\cal O}(v^2/c^2)
\end{equation}
\begin{equation}
m_1\dot{y}_1= +\frac{\hbar}{r^2}\left [ x-\frac{2x}{rc}(x\dot{x}_2+y\dot{y}_2)+\frac{r}{c}\dot{x}_2\right ] +{\cal O}(v^2/c^2)
\end{equation}
\begin{equation}
m_2\dot{y}_2= -\frac{\hbar}{r^2}\left [ x+\frac{2x}{rc}(x\dot{x}_1+y\dot{y}_1)-\frac{r}{c}\dot{x}_1\right ]  +{\cal O}(v^2/c^2)
\end{equation}
So that the components of the velocity of the center of mass is given by:
\begin{equation}
(m_1+m_
2)\dot{X}_{c.m.}= +\frac{2\hbar}{c}\frac{y}{r^3}\left [ x(\dot{x}_1+\dot{x}_2)+y(\dot{y}_1+\dot{y}_2)\right ]-\frac{\hbar}{cr}(\dot{y}_1+\dot{y}_2)+{\cal O}(v^2/c^2)
\end{equation}
\begin{equation}
(m_1+m_2)\dot{Y}_{c.m.}= -\frac{2\hbar}{c}\frac{x}{r^3}\left [ x(\dot{x}_1+\dot{x}_2)+y(\dot{y}_1+\dot{y}_2)\right ]+\frac{\hbar}{cr}(\dot{x}_1+\dot{x}_2)+{\cal O}(v^2/c^2)
\end{equation}
which shows that the center of mass velocity is not zero, as is predicted
by NBT. The center of mass is self-accelerated
.
\par
An estimation of the center of mass velocity is simple. From the above equations we
have:
\[ v_1\ \ or \ \ v_2 \sim \frac{\hbar}{mr} \]
\[ v_{c.m.} \sim \frac{\hbar ^2}{m^2c}\frac{1}{r^2} \sim \frac{10^{-68}}{10^{-60}\times 10^8}\frac{1}{10^{-20}} \sim 10^4\ m/sec \sim 10^{-4}c \]
which must be an observable effect. If one considers a gas of such atoms and 
assumes that this self-acceleration  is converted to heat via collisions, one
has:
\[ \frac{1}{2}mv_{c.m.}^2\sim \frac{3}{2}kT \Longrightarrow T\sim 10 ^{\circ}K\]
Thus, this effect can be observed as the self-heating of a gas of atoms 
having a temperature of about
ten degrees of Kelvin!!
\section{Unstability of Hydrogen-like Atoms}
\par
Now, we want to investigate to what extent the motion of a Hydrogen-like atom differs
from the standard one, i.e. the one predicted by NBT. For simplicity we assume 
that $m_1=m_2=m$. The equations (16)--(19) can be written in the spherical coordinates 
in the following form (we assume that $\theta=\pi/2$):
\begin{equation}
\dot{X}_{c.m.}=\dot{Y}_{c.m.}=0
\end{equation}
\begin{equation}
\dot{r}=-\frac{2c}{1-r^2/\alpha^2}
\end{equation}
\begin{equation}
\dot{\phi}=\frac{(2\alpha c/r)\cos \phi-\dot{r}(\sin \phi -(\alpha/r)\cos \phi)}{r\cos \phi -\alpha \sin \phi}
\end{equation}
where:
\begin{equation}
\alpha=\frac{\hbar}{mc}
\end{equation}
The solutions of (23) and (24) are:
\begin{equation}
\frac{r^3}{3\alpha^2}-r=\frac{r_0^3}{3\alpha^2}-r_0+2ct 
\end{equation}
\begin{equation}
r=r
_0+\alpha \phi
\end{equation}
where $r_0=r(t=0)$. Note that as $c\rightarrow \infty$ these are the same as 
the results of NBT. A glance at these relations leads to the strange result
that this sort of atom is not stable, in the sense that $r$ is an increasing function of 
time. In fact $r$ increases from $1\times 10^{-10}$ meter to $1$ meter in the
time interval:
\[ t\sim \frac{m^2c}{6\hbar^2}\sim10^{16}\ sec\ \sim 10^9\ year\ \sim 10^{-1}\ times\ of\ the\ age\ of\ the\ universe.\]
\section{The Dis
tribution Function}
\par
Since we calculate the motion of particles via (4), rather than (2), there is no 
nessecity for (3) to be consistent. In fact, in order to have a consistent theory, 
one must calculate the distribution function via the conservation law of particles,
not using (3). The conservation law can be written as:
\begin{equation}
\frac{\partial \rho}{\partial t}+\vec{\nabla}\cdot (\rho \vec{v})=0
\end{equation}
or equivalently as
\begin{equation}
\frac{\partial \ln \rho}{\partial t
}+\vec{v}\cdot\vec{\nabla}(\ln \rho)+\vec{\nabla}\cdot \vec{v}=0
\end{equation}
To solve this equation for the system presented in the previous section,
perturbatively, we write:
\begin{equation}
\ln \rho =\ln \rho _0(r,\theta)+\alpha \eta(r,\theta,\phi;t)+\cdots
\end{equation}
where $\rho_0$ is the result of NBT. Equation (29) leads to the following result, for
first order terms:
\begin{equation}
\frac{\partial \eta}{\partial t}+\frac{2\alpha c}{r^2}\frac{\partial (\ln \rho_0)}{\partial r}+\frac{
2\alpha c}{r^2}\frac{\partial \eta}{\partial \phi}=0
\end{equation}
with the solution:
\begin{equation}
\eta(r,\theta;t)=-\frac{2\alpha c}{r^2}\frac{\partial (\ln \rho_0)}{\partial r}t
\end{equation}
So that:
\begin{equation}
\rho=\rho_0 e^{-(2\alpha c/r^2)[\partial (\ln \rho_0)/\partial r]\alpha t +{\cal O}(\alpha^2)}
\end{equation}
which shows the unstability of Hydrogen-like atoms clearly. Note that we have
not normalised $\rho$, because this must be done when all orders are calculated.
\par

This is again an observable effect, because change in the distribution function
would be reflected in physical quantities like electric dipole moment, magnetic
quadrapole moment, etc.
\section{The Problem of Energy}
\par
As it is stated by Squires[4], RBT is ambigous when the $\psi$-field is an
explicit function of time. This is because we do not know what time to
use in (4). One way out of
this problem is to postulate that such explicit time dendencies should be evaluated at time
$t_{ii}=t$.
\par
In NBT, the energy of the system is:
\begin{equation}
{\cal E}=i\hbar\left [ \frac{\partial}{\partial t}\left \{ \ln \left ( \frac{\psi}{\sqrt{\psi^*\psi}}\right ) \right \} \right ]_{\{\vec{x}_j\}=\{\vec{r}_j(t)\}}
\end{equation}
which can be shown to be equivalant to:
\begin{equation}
{\cal E}=\frac{1}{2}\sum_{i=1}^N m_i|\dot{\vec{r}}_i(t)|^2+[V(\vec{x}_1,\cdots, \vec{x}_N;t)+Q(\vec{x}_1,\cdots,\vec{x}_N;t)]_{\{\vec{x}_j\}=\{\vec{r}_j(t)\}}
\end{equation}
where the quantum potential is given 
by:
\begin{equation}
Q(\vec{x}_1,\cdots,\vec{x}_N;t)=\sum_{i=1}^N -\frac{\hbar^2}{2m_i}\frac{\nabla_i^2\sqrt{\psi^*\psi}}{\sqrt{\psi^*\psi}}
\end{equation}
The extension of (34) and (35) to RBT is ambigous. But if $\psi$ has a time dependence
like $e^{-iEt/\hbar}$ then (34) works and leads to ${\cal E}=E$.
\par
It seems to us that there is a natural solution to this problem. First, we note that 
the classical potential $V$ in (35) is in fact a retarded potential. That is, for
the system in section 
3, it is equal to:
\begin{equation}
V(\vec{x}_1,\vec{x}_2;t)=q_1\varphi(\vec{x}_1;t)+q_2\varphi(\vec{x}_2;t)-q_1\vec{A}(\vec{x}_1;t)\cdot\dot{\vec{r}}_1-q_2\vec{A}(\vec{x}_2;t)\cdot\dot{\vec{r}}_2
\end{equation}
where $\varphi$ and $\vec{A}$ are the retarded electromagnetic potentials and 
$q_1$ and $q_2$ are the charges of two particles. Clearly in the calculation of 
${\cal E}$ we must use $V(\vec{r}_1(t),\vec{r}_2(t);t)$. 
\par
On the other hand, the $i$th term in the sum in (36) represents the $
i$th particle
contribution to the quantum potential. So, we define:
\begin{equation}
\tilde{Q}=\sum_{i=1}^N \left [ -\frac{\hbar^2}{2m_i}\frac{\nabla_i^2\sqrt{\psi^*\psi}}{\sqrt{\psi^*\psi}}\right]_{\{\vec{x}_j\}=\{\vec{r}_j(t_{ij})\}}
\end{equation}
and assume that the correct energy of the system is given by:
\begin{equation}
{\cal E}=\frac{1}{2}\sum_{i=1}^N m_i[\dot{\vec{r}}_i(t)]^2+V(\vec{r}_1(t),\cdots, \vec{r}_N(t);t)+\tilde{Q}
\end{equation}
It is obvious that this is very different from $E$
 of the example in section 3.
\section{References}
\par
[1]-- D. Bohm, Phys. Rev. {\bf 85} (1952) 166, 180.
\newline
[2]-- P.R. Holland, {\it The quantum theory of motion\/}, (1993) Cambridge University Press, London/New York.
\newline
[3]-- D. Bohm, B.J. Hiley, and P.V. Kaloyerou, Phys. Rep. {\bf 144}, 6 (1987) 321, 349.
\newline
[4]-- E.J. Squires, Phys. Lett. {\bf A 178} (1993) 22.
\newline
[5]-- S. Mackman and E.J. Squires, Found. Phys. {\bf 25} (1995) 391.
\newline
\end{bf}
\end{document}